\documentclass[modern]{aastex61}

\usepackage{graphicx}
\usepackage{amsmath}
\usepackage{amssymb}
\usepackage{enumitem}

\newcommand\mybar{\kern1pt\rule[-\dp\strutbox]{.8pt}{\baselineskip}\kern1pt}

\setlist[itemize]{noitemsep, topsep=0pt, leftmargin=*}

\shorttitle{Evolution of the CMB Quadrupole}
\shortauthors{Loeb}

\begin{document}

\title{Time Evolution of the CMB Quadrupole}

\author{Abraham Loeb}
\affiliation{Astronomy Department, Harvard University, 60 Garden
  St., Cambridge, MA 02138, USA}

\begin{abstract}
I show that the quadrupole of the Cosmic Microwave Background (CMB)
evolves more rapidly than previously expected, as a result of the
acceleration of the Sun towards the Galactic center. The acceleration,
measured most recently by {\it Gaia} EDR3, implies a fractional change
in the quadrupole of $\sim 10^{-9}$ per year, an order of magnitude
larger than expected from the evolution in the last scattering surface
of the CMB.

\end{abstract}

*\section*{}
\section*{Introduction}
Within a timescale of order the age of the Universe, $\sim 10^{10}$~yr,
the large-scale anisotropies of the Cosmic Microwave Background (CMB)
are expected to change as a result of the change in the location of
the last scaterring surface \citep{2007ApJ...671.1075L,2007PhRvD..76l3010Z,2008PhRvD..77d3505M}. Naively, one
would expect anisotropies on the largest scales - such as the
quadrupole moment, to change most slowly. However, here we show this
not to be true as a result of the acceleration of the Sun towards the
center of the Milky Way. This acceleration was measured most recently
by {\it Gaia} EDR3 to be, ${\dot v}=2.32(\pm0.16)\times 10^{-8}~{\rm
  cm~s^{-2}}$, towards the Galactic center
\citep{2021A&A...649A...9G}.

\section*{Calculation}

The velocity of the Sun relative to the cosmic frame of reference,
${\vec {\bf v}}$, results in a CMB dipole moment, which is routinely
removed from CMB anistropy maps. But to second-order in ${\vec {\bf
    \beta}}\equiv ({\vec {\bf v}}/c)$, the motion also leads to a
quadrupole anisotropy with an angular dependence of $(\cos^2\theta
-1/3)$, where $\theta$ is the angle between the velocity vector ${\vec
  {\bf v}}$ and the photon direction.

The fractional change in the CMB intensity as a result of the
kinematic quadrupole depends on photon frequency
\citep{2003PhRvD..67f3001K},
\begin{equation}
\left({\Delta I_\nu\over I_\nu}\right)_Q=F(x) \beta^2 ,
\label{one}
\end{equation}
with $F(x)=[xe^x/(e^x-1)][(x/2)\coth(x/2)]$. Here, $x=(h\nu/kT)$ with
$I_\nu=[2(kT)^3/(hc)^2] x^3/(e^x-1)$ being the CMB blackbody intensity
at a frequency $\nu$ and the mean CMB temperature $T=2.725$K
\citep{2009ApJ...707..916F}. At the low frequencies of the
Rayleigh-Jeans regime where $x\ll1$, we get $F(x)\approx 1$ and
$(\Delta I_\nu/I_\nu)=(\Delta T/T)$.

\section*{Conclusions}

The time derivative of equation (\ref{one}) yields a kinematic
variation in the quadrupole, which for $x\ll1$ is given by,
\begin{equation}
\left({{\dot T}\over T}\right)_Q= 2{\vec {\bf \beta}}\cdot{\dot {\vec {\bf \beta}}} ,
\label{two}
\end{equation}
where from the {\it Gaia} measurement, ${\dot \beta}=2.44\times
10^{-11}~{\rm yr^{-1}}$, and from the CMB dipole $\beta \approx
1.23\times 10^{-3}$
\citep{1993ApJ...419....1K,1994ApJ...420..445F}. The dot-product of
the acceleration vector towards the Galactic center and the CMB dipole
velocity vector yields a geometric projection factor of $0.065$
\citep{2008MNRAS.386.2221L}.

Given the quadrupole moment measured by the Planck satellite, $Q\equiv
(\Delta T/T)_Q\sim 4.5\times 10^{-6}$ \citep{2015JCAP...06..047N}, the
fractional time derivative of the CMB quadrupole, as a result of the
acceleration of the Sun towards the Galactic center, is given by,
\begin{equation}
\left({{\dot Q}\over Q}\right) \approx 10^{-9}~{\rm yr^{-1}} .
\label{three}
\end{equation}
This rate of evolution is over an order of magnitude larger than
expected from the change in the last scattering surface of the CMB
\citep{2007ApJ...671.1075L,2007PhRvD..76l3010Z,2008PhRvD..77d3505M}.

The expected time evolution of the CMB quadrupole spectrum, $F(x)$, in
equation (\ref{one}), could potentially be measured - especially for
$x\gg 1$ where $F(x)\gg 1$, with future CMB spectral distortion
experiments \citep{2021ExA....51.1515C}.

\bigskip
\bigskip
\section*{Acknowledgements}

This work was supported in part by Harvard's {\it Black Hole
  Initiative}, which is funded by grants from JFT and GBMF.

\bibliographystyle{aasjournal}
\bibliography{Q}
\label{lastpage}
\end{document}